\pdfoutput=1
\documentclass[11pt,twoside,a4paper,cmspaper,final,collab]{cms-tdr}
\def\svnVersion{ab18241}\def\svnDate{2023/09/19}\def\cmsCernNoTag{CERN-EP-2022-248}\def\cmsCernDate{\today}\def\cmsMessage{Published in Physical Review Letters as \href{http://dx.doi.org/10.1103/PhysRevLett.131.121901}{\doi{10.1103/PhysRevLett.131.121901}.}}
\begin{document}\cmsNoteHeader{BPH-21-001}

\ifthenelse{\boolean{cms@external}}{\providecommand{\suppMaterial}{\cite{suppMatBib}}}{\providecommand{\suppMaterial}{Appendix~\ref{app:suppMat}}}
\newlength{\cmsFigWidth}
\ifthenelse{\boolean{cms@external}}{\setlength\cmsFigWidth{\columnwidth}}{\setlength\cmsFigWidth{0.65\textwidth}}

\ifthenelse{\boolean{cms@external}}{\providecommand{\cmsLeft}{upper\xspace}}{\providecommand{\cmsLeft}{left\xspace}}
\ifthenelse{\boolean{cms@external}}{\providecommand{\cmsRight}{lower\xspace}}{\providecommand{\cmsRight}{right\xspace}}

\ifthenelse{\boolean{cms@external}}{\providecommand{\cmseqdos}{ Eq.~(2)}}{\providecommand{\cmseqdos}{ Eq.~(\ref{eq:fdoverfuDefinition})}}
\ifthenelse{\boolean{cms@external}}{\providecommand{\cmsfigone}{ Fig.~2}}{\providecommand{\cmsfigone}{ Fig.~\ref{fig:Rs}}}
\ifthenelse{\boolean{cms@external}}{\providecommand{\cmsfigtwo}{ Fig.~3}}{\providecommand{\cmsfigtwo}{ Fig.~\ref{fig:fdfu}}}

\cmsNoteHeader{BPH-21-001}

\newcommand{\Bu}{\ensuremath{\PBp}\xspace}
\newcommand{\Bs}{\ensuremath{\PBzs}\xspace}
\newcommand{\Bd}{\ensuremath{\PBz}\xspace}
\newcommand{\Bc}{\ensuremath{\PBpc}\xspace}
\newcommand{\LambdaB}{\ensuremath{\PGLzb}\xspace}

\newcommand{\BuJpsiK}{\ensuremath{\PBp \! \to \! \PJGy \, \PK^{+}}\xspace}
\newcommand{\BsJpsiPhi}{\ensuremath{\PBzs \! \to \! \PJGy \, \PGf}\xspace}
\newcommand{\BdJpsiKstar}{\ensuremath{\PBz \! \to \! \PJGy \, \PKst{}^{0}}\xspace}

\newcommand{\Jpsik}{\ensuremath{\PJGy \, \PKp}\xspace}
\newcommand{\JpsiPhi}{\ensuremath{\PJGy \, \PGf}\xspace}
\newcommand{\JpsiKstar}{\ensuremath{\PJGy \, \PKst{}^{0}}\xspace}

\newcommand{\Phikk}{\ensuremath{\PGf \! \to \! \PKp \PKm}\xspace}
\newcommand{\Kpik}{\ensuremath{\PKst{}^{0} \! \to \! \Pgpm \PKp}\xspace}

\newcommand{\fu}{\ensuremath{f_\PQu}\xspace}
\newcommand{\fd}{\ensuremath{f_\PQd}\xspace}
\newcommand{\fs}{\ensuremath{f_\PQs}\xspace}

\newcommand{\fsfu}{\ensuremath{f_\PQs/f_\PQu}\xspace}
\newcommand{\fdfu}{\ensuremath{f_\PQd/f_\PQu}\xspace}

\newcommand{\fsfd}{\ensuremath{f_\PQs/f_\PQd}\xspace}
\newcommand{\Rs}{\ensuremath{{\mathcal{R}_\PQs} }\xspace}
\newcommand{\Rd}{\ensuremath{{\mathcal{R}_\PQd} }\xspace}
\newcommand{\Rpmz}{\ensuremath{{\mathcal{R}^{\pm,0} }}\xspace}

\title{Measurement of the dependence of the hadron production fraction ratios
\texorpdfstring{\fsfu and \fdfu on \PB meson kinematic variables in proton-proton collisions at $\sqrt{s} = 13$\,TeV}
{fs/fu and fd/fu on B meson kinematic variables in proton-proton collisions at sqrt(s)=13TeV}}

\author*[cern]{The CMS Collaboration}

\date{\today}

\abstract{The dependence of the ratio between the \Bs and \Bu hadron production fractions, \fsfu, 
on the transverse momentum (\pt) and rapidity of the \PB mesons 
is studied using the decay channels \BsJpsiPhi and \BuJpsiK. 
The analysis uses a data sample of proton-proton collisions at a center-of-mass energy of 13\TeV, 
collected by the CMS experiment in 2018 and corresponding to an integrated luminosity of 61.6\fbinv. 
The \fsfu ratio is observed to depend on the \PB \pt 
and to be consistent with becoming asymptotically constant at large \pt. 
No rapidity dependence is observed. 
The ratio of the \Bd to \Bu meson 
production fractions, \fdfu, 
is also measured, for the first time in proton-proton collisions,
using the \BdJpsiKstar decay channel. 
The result is found to be within one standard deviation of unity 
and independent of \pt and rapidity, as expected from isospin invariance.}

\hypersetup{%
pdfauthor={CMS Collaboration},%
pdftitle={Measurement of the dependence of the hadron production fraction ratios fs/fu and fd/fu on B meson kinematic variables in proton-proton collisions at sqrt(s) = 13 TeV},%
pdfsubject={CMS},%
pdfkeywords={CMS, B physics, rare B decays, B meson production}}

\maketitle

When a \PQb quark is produced with sufficient momentum, 
it undergoes a process, referred to as fragmentation or hadronization, 
that results in the creation of a \PQb hadron. 
The relative rates of the various flavors of \PQb hadrons are referred to as hadron production fractions 
or fragmentation functions. 
The fractions of \Bu, \Bd, and \Bs mesons are denoted by \fu, \fd, and \fs, respectively.
Experiments at the \PB factories use the characteristics of \PB meson production
at the \PgUd to set the scale of the measured \Bu and \Bd branching fractions,
allowing precision measurements at both \PB factories and hadron colliders.
For the \Bs, however, the limited event samples at the \PgUe and uncertainties
in the fraction of \Bs mesons produced do not allow the same approach.
As a result, precision \Bs branching fraction measurements at hadron colliders
rely on ratios to \Bu or \Bd decay modes,
which require knowledge of \fsfu or \fsfd, respectively. 
In fact, measurements of the rare decay $\Bs \to \MM$, 
which is used to search for physics beyond the standard model, 
are currently limited by the uncertainty in 
\fsfu~\cite{CMS:2014xfa,Aaboud:2018mst,CMS:2019bbr,LHCb:2017rmj,LHCb:2021awg}. 
Besides providing information about the nature of the strong interaction,
precise \fsfu values are also needed to measure the $\Bs \to \MM$ 
and other \Bs branching fractions.

A variety of measurements obtained from \PZ boson decays 
by the LEP experiments have been combined by the Heavy Flavor Averaging Group 
to obtain values for the hadron production fractions and the ratio \fsfd~\cite{HFAG2018}. 
Subsequent central-rapidity measurements 
by CDF (for pseudorapidity $\abs{\eta} < 1$) in proton-antiproton collisions at $\sqrt{s} = 1.96$\TeV~\cite{CDF} and 
by ATLAS (for $\abs{\eta} < 2.5$) in proton-proton ($\Pp\Pp$) collisions at $\sqrt{s} = 7$\TeV~\cite{atlas_2015} 
were compatible with the LEP result, with no evidence of a dependence on transverse momentum (\pt). 
Instead, measurements of \fsfd in the forward-rapidity region ($2 < \eta < 6.4$) 
by LHCb in $\Pp\Pp$ collisions at $\sqrt{s} = 7, 8$, and $13$\TeV~\cite{2103.06810} 
(combining results from Refs.~\cite{1111.2357, 1301.5286, 1910.09934, 1902.06794}) 
give strong evidence for a \pt dependence of these quantities, 
with the values decreasing as the \pt of the \PB meson increases.

This Letter presents an analysis aimed at establishing
if and how the relative \Bs and \Bu production rates change with \pt
in a kinematic region relevant for
the ATLAS and CMS experiments at the CERN LHC,
$\pt > 12$\GeV and $\abs{y}<2.4$,
approximately complementary to that of the LHCb detector. 
Additionally, we perform the first test of the isospin invariance 
in \PB meson production in $\Pp\Pp$ collisions, by measuring the \fdfu ratio.
These measurements use a sample of $\Pp\Pp$ collisions
collected by the CMS experiment in 2018 at a center-of-mass energy of 13\TeV and
corresponding to an integrated luminosity of 61.6\fbinv~\cite{CMS-PAS-LUM-2018,CMS:LUM-17-003}.

Throughout this Letter, charge-conjugate states are implicitly included,
and $\PKst{}^{0}$ and \PGf represent the $\PKst{}$(892)$^{0}$ and \PGfP{1020}, respectively.
The \Bu and \Bs mesons are reconstructed using the \BuJpsiK and \BsJpsiPhi decay channels,
with the \PJGy and \PGf mesons decaying as $\PJGy \to \MM$ and \Phikk.
The respective event yields, $N_{\Bu}$ and $N_{\Bs}$, are measured
with corresponding detection efficiencies $\epsilon_{\Bu}$ and $\epsilon_{\Bs}$.
The ratio of the efficiency-corrected meson yields,
$\Rs = (N_{\Bs} / \epsilon_{\Bs}) \, / \, (N_{\Bu} / \epsilon_{\Bu})$,
is directly proportional to the \fsfu ratio,
\begin{linenomath}
\begin{equation}
\Rs = \fsfu \; \frac{\mathcal{B}(\BsJpsiPhi) \mathcal{B}(\Phikk)} {\mathcal{B}(\BuJpsiK)},
\label{eq:fsoverfuDefinition}
\end{equation}
\end{linenomath}
where $\mathcal{B}(\BsJpsiPhi)$, $\mathcal{B}(\Phikk)$, and $\mathcal{B}(\BuJpsiK)$
are the branching fractions of the indicated decay channels;
the $\mathcal{B}(\PJGy \to \MM)$ factor cancels in the ratio.
Given that the available measurements of the \BsJpsiPhi branching fraction
and of \fs are correlated, we report measurements of \Rs rather than of \fsfu.

The analysis also includes a measurement of the ratio
between the \Bd and \Bu hadron fractions, \fdfu,
using the \Bd yield determined with \BdJpsiKstar events,
where the $\PKst{}^{0}$ mesons are reconstructed in the \Kpik decay channel.
Using notations analogous to those used above,
\begin{linenomath}
\begin{equation}
\ifthenelse{\boolean{cms@external}}{
\begin{split}
\Rd & = \frac{N_{\Bd}}{\epsilon_{\Bd}} \, \Bigg/ \, \frac{N_{\Bu}}{\epsilon_{\Bu}} \\
      & =  \fdfu \; \frac{\mathcal{B}(\BdJpsiKstar) \mathcal{B}(\Kpik) }{ \mathcal{B}(\BuJpsiK) }.
\end{split}
}{
\Rd = \frac{N_{\Bd}}{\epsilon_{\Bd}} \, \Bigg/ \, \frac{N_{\Bu}}{\epsilon_{\Bu}}
 = \fdfu \; \frac{\mathcal{B}(\BdJpsiKstar) \mathcal{B}(\Kpik) }{\mathcal{B}(\BuJpsiK)}.
}
\label{eq:fdoverfuDefinition}
\end{equation}
\end{linenomath}

Under the assumption of strong isospin symmetry,
the \fdfu ratio is expected 
to be independent of kinematic variables and identical to unity.
Given that the branching fractions used to calculate \fdfu from \Rd
are (independently) obtained from high-precision 
\PB factory measurements,
we report the more relevant quantity \fdfu. 

The CMS apparatus is a multipurpose detector~\cite{CMS:2008xjf}
designed to trigger on and identify electrons, muons, photons,
and (charged and neutral) hadrons~\cite{CMS:2020uim,CMS:2018rym,CMS:2014pgm}.
A superconducting solenoid of 6\unit{m} internal diameter provides a magnetic field of 3.8\unit{T}.
Within the solenoid volume are the silicon pixel and strip tracker,
a crystal electromagnetic calorimeter, and a brass and scintillator hadron calorimeter.
Muons are measured in gas-ionization detectors embedded in the steel flux-return yoke outside the solenoid.
Events of interest are selected using a two-tiered trigger system.
The first level, composed of custom hardware processors,
uses information from the calorimeters and muon detectors to select events
at a rate of 100\unit{kHz} within a fixed latency of about 4\mus~\cite{CMS:2020cmk}.
The second level,
consisting of a farm of processors running a faster version of the full event reconstruction software,
reduces the rate to around 1\unit{kHz}, before data storage~\cite{CMS:2016ngn}.

The events used in the analysis were selected by a trigger requiring
two oppositely charged muons with $\abs{\eta} < 2.5$ and  $\pt > 4$\GeV,
a distance of closest approach between the two muons smaller than 0.5\cm,
a dimuon vertex fit $\chi^2$ probability larger than 10\%,
a dimuon invariant mass in the range 2.9--3.3\GeV,
and a  transverse distance between the dimuon vertex and the beam axis, $L_{xy}$, larger than three times its uncertainty.
In addition, the dimuon $\vec{p}_\mathrm{T}$ and transverse displacement vector, $\vec{L}_{xy}$,
must be aligned with each other:
${\cos\theta} \equiv {\vec{L}_{xy} \cdot \vec{p}_\mathrm{T} / (L_{xy} \, \pt)} > 0.9$.
The trigger also requires a third track in the event,
compatible with being produced at the dimuon vertex, having $\pt > 1.2$\GeV
and a transverse impact parameter significance larger than 2.
Finally, the dimuon-plus-track vertex fit $\chi^2$ probability must be larger than 10\%.

The charged tracks used to reconstruct the \PB mesons must pass high-purity criteria~\cite{CMS:2014pgm},
have five or more hits in the silicon tracker, at least one of them in the pixel layers,
and have $\abs{\eta} < 2.4$.
They must also match the tracks that triggered the data readout.
The muons used to reconstruct the \PJGy candidates must fulfill
soft-muon identification requirements~\cite{CMS:2018rym},
which include the (loose) matching between the track reconstructed in the silicon tracker
and the one reconstructed in the muon detectors.
They must also have $\pt > 4$\GeV and an impact parameter
smaller than 0.3\cm in the transverse plane and smaller than 20\cm along the beam axis.

The \PJGy candidate is combined with one track
to reconstruct \BuJpsiK decays
or with a pair of oppositely charged tracks
to reconstruct the \BsJpsiPhi and \BdJpsiKstar decays.
All three tracks must have $\pt > 1.2$\GeV.
They are fitted together with the dimuon, imposing a common (secondary) vertex
(SV, the \PB meson decay point),
constraining the dimuon invariant mass to the \PJGy world-average mass,
$m_\PJGy^\mathrm{PDG}$~\cite{PDG2022},
and assigning to each of the other tracks the $\Pgppm$ or $\PKpm$ masses, as suitable.
Furthermore, the invariant mass of the pair of tracks must satisfy
$\abs{M(\PKp \PKm)-m_{\PGf}^{\mathrm{PDG}}} < 10$\MeV for \Bs decays and
$\abs{M(\Pgp \PK)-m_{\PKst^{0}}^{\mathrm{PDG}}} < 50$\MeV for \Bd decays,
where $m_{\PGf}^{\mathrm{PDG}}$ and $m_{\PKst^{0}}^{\mathrm{PDG}}$ are,
respectively, the $\PGf$ and $\PKst^{0}$ world-average masses~\cite{PDG2022}.
In the latter case, lack of particle identification
implies that both combinations are possible ($\Pgpp \PKm$ and $\PKp \Pgpm$);
if both match the mass window requirement, only the one with mass closer to
$m_{\PKst^0}^\mathrm{PDG}$ is kept.

The primary vertex (PV) is selected among the several reconstructed $\Pp\Pp$ collisions
as the one that minimizes the pointing angle of the \PB meson,
defined as the angle between the \PB momentum
and the vector joining the primary and secondary vertices.
The PV is refitted without the tracks of the \PB candidate
before computing the \PB decay length as the distance between the PV and the SV.
We select \PB meson candidates with $12 < \pt < 70$\GeV, $\abs{y}<2.4$,
a decay length larger than five times its uncertainty,
and a dimuon-plus-track(s) vertex $\chi^2$ probability larger than 10\%.
For each decay channel, if more than one \PB meson candidate is reconstructed in an event
(occurring in less than 1\% of the events),
only the one with the highest fit $\chi^{2}$ probability is kept.

The event selection criteria described above were optimized through the study of
Monte Carlo (MC) event samples,
which were also used to evaluate the detection efficiencies
and the shapes of the invariant mass distributions of some background contributions.
They were generated with \PYTHIA\,8.230~\cite{Pythia} for the production and hadronization steps,
with \EVTGEN\,1.6.0~\cite{EvtGen} for the decay of the \PQb hadrons,
and with \PHOTOS\,3.61~\cite{PHOTOS2} for the final-state radiation modeling.
The response of the CMS detector to the generated events,
including the trigger and reconstruction steps,
was simulated with \GEANTfour~\cite{Geant4},
using algorithms identical to those used on the data.
The simulated events include multiple $\Pp\Pp$ interactions in the same or nearby beam crossings,
with a distribution matching the one observed in the collected data.

\begin{figure*}[ht]
\centering
\includegraphics[width=\textwidth]{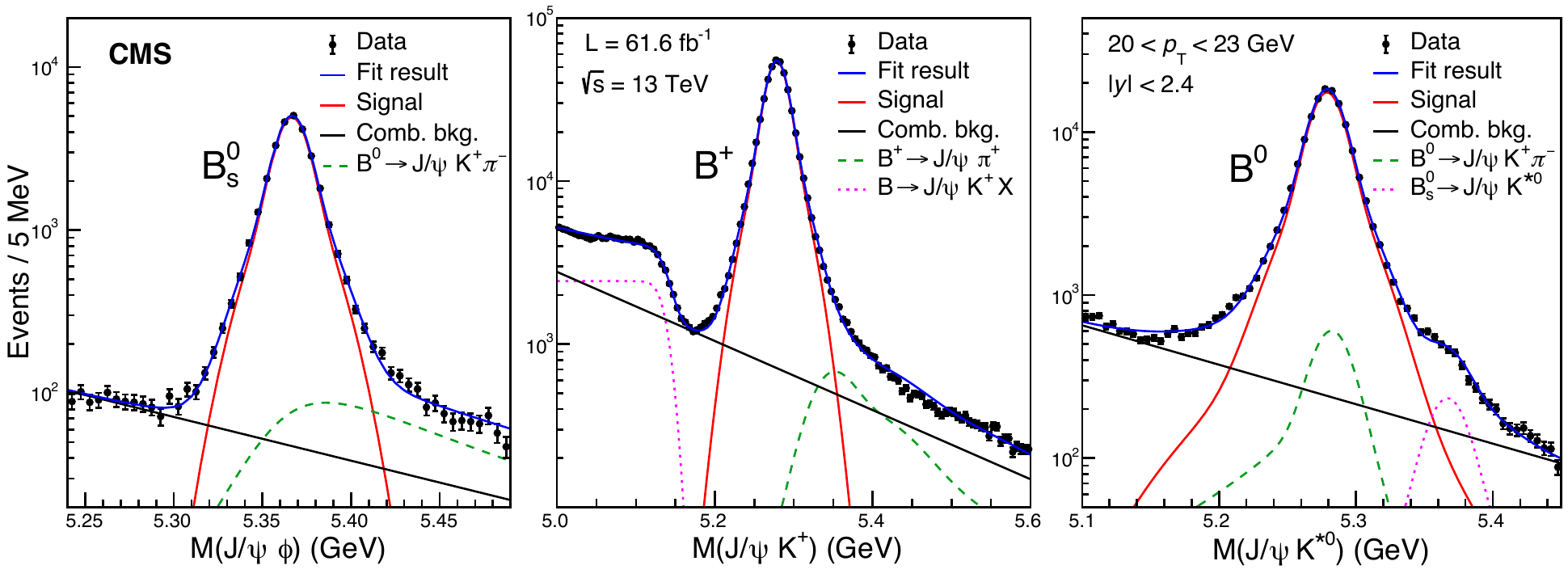}
\caption{The \JpsiPhi, \Jpsik, and \JpsiKstar invariant mass distributions
for \PB meson candidates with $20< \pt < 23$\GeV,
and associated fits as described in the text.}
\label{fig:mass-distributions}
\end{figure*}

The \Bs, \Bu, and \Bd meson yields are measured by fitting,
with unbinned maximum likelihood techniques,
the \JpsiPhi, \Jpsik, and \JpsiKstar invariant mass distributions, respectively.
These distributions are fitted for 12 \pt bins (integrated over $y$)
or 7 $\abs{y}$ bins (integrated over \pt),
with ranges defined so as to keep a similar number of events in each bin.
Figure~\ref{fig:mass-distributions} shows the three invariant mass distributions for
the 20--23\GeV \pt bin.

The signal peak is fitted by the sum of two Gaussian functions
with a common mean and independent widths,
reflecting the shapes observed in the simulated event samples;
the mean and widths are left free in the fit.
The underlying combinatorial background is fitted by an exponential function.
The \Jpsik sample includes a background term due to events
where \PB mesons decay through $\PJGy \, \PK^{+} \, X$ channels
and the $X$ particle is not reconstructed.
This contribution is described by an error function,
with two free shape parameters.
Some of the \JpsiKstar candidates have swapped pion-kaon mass assignments.
They are included in the fit model by adding a component
with shape and normalization (12\% relative to the unswapped yield) fixed from simulation.
The \JpsiKstar sample also includes (Cabibbo-suppressed) $\Bs \! \to \! \PJGy \, \PKst^{0}$ decays,
with normalization as a free parameter
and described by a shape identical to that of the \Bd signal peak,
except for the shifted mass and for a small width broadening to account for the change of
mass resolution, as determined from simulation.

All the other background contributions
have shapes determined from the simulated event samples.
The $\Bu \! \to \! \PJGy \, \Pgpp$ curve in the \Bu panel represents
(Cabibbo-suppressed) decays where the pion track is misinterpreted as a kaon;
its normalization is fixed to that of the \Bu signal yield,
scaled by the ratio of the two branching fractions~\cite{PDG2022}.
The misidentification of a pion as a kaon is also the reason why sometimes
a $\Bd \! \to \! \PJGy \, \PK \, \Pgp$ decay is incorrectly assigned to the \Bs sample.
This small contribution is described by a Johnson function~\cite{Johnson},
with a normalization constrained, in each \pt or $\abs{y}$ bin, to that of the \Bs signal yield,
the scaling factor being the relative yield found in a fit to the integrated event sample.
The background to the \Bs sample from
$\PGLzb \! \to \! \PJGy \, \PKm \, \Pp$ decays where the proton is misidentified as a kaon
has been found to be negligible.
A background from $\Bd \! \to \! \PJGy \, \PK \, \Pgp$ decays also contributes to the \BdJpsiKstar distributions.
This peaking background is modeled with a double-sided Crystal Ball~\cite{bib:CB} function plus a Gaussian function.
Its normalization is constrained, in each \pt or $\abs{y}$ bin, to that of the \Bd signal yield,
scaled by the yield ratio obtained in the fit of the integrated event sample.

As seen in Eqs.~(\ref{eq:fsoverfuDefinition}) and~(\ref{eq:fdoverfuDefinition}),
only the ratios of detection efficiencies,
$\epsilon_{\Bs} \, / \, \epsilon_{\Bu}$ and
$\epsilon_{\Bd} \, / \, \epsilon_{\Bu}$,
are needed to convert the ratios of signal event yields,
obtained from the fits illustrated in Fig.~\ref{fig:mass-distributions},
into the \Rs and \Rd observables.
These efficiency ratios are evaluated using the simulated event samples,
reflecting the trigger and reconstruction steps, as well as the detector acceptance.
Both ratios increase by around a factor of 3.5 between the lowest and highest \pt bins,
while the variation with $\abs{y}$ is only at the 10\% level.

The \Rs and \Rd measurements are affected
by systematic uncertainties in the determination of the fitted signal yields
and in the evaluation of the efficiency ratios.

The systematic uncertainties affecting the signal yields are evaluated by
repeating the fits of the mass distributions in alternative conditions and computing
the difference between the obtained results and those of the baseline fit.
Two main variations of the fit model are independently considered:
first, the modeling of the signal peaks is changed from the default double-Gaussian
to a Student's t-distribution~\cite{Studentt};
second, the combinatorial background is fitted by a first-order Chebyshev polynomial
instead of the baseline exponential function.
An additional systematic uncertainty in the \Bd meson yield,
of less than 1\%, is evaluated by fitting the \JpsiKstar mass distribution
changing the normalization of the ``$\pi$--K swap" term,
relative to that of the \Bd signal term,
by the uncertainty in the default value,
which exclusively reflects the sizes of the MC event samples;
other systematic effects were found to be negligible.
The fit procedure itself is seen to provide unbiased results,
for each of the bins,
both for the central values and the uncertainties,
through a study involving fits of 1000 event samples randomly generated
using the nominal functions with the best fit parameters
and with sizes corresponding to the number of measured events.
The result is that the uncertainties in the fitted \Bs, \Bu, and \Bd signal yields
contribute systematic uncertainties to the \Rs and \Rd measurements that
are, respectively, in the 1.6--2.6\% and 2.0--5.0\% ranges.

For the efficiency ratios,
$\epsilon_{\Bs} \, / \, \epsilon_{\Bu}$ and $\epsilon_{\Bd} \, / \, \epsilon_{\Bu}$,
a systematic uncertainty, $\approx$1\% for all \pt and $\abs{y}$ bins,
reflects the size of the simulated event samples.
As the \Bs and \Bd decays lead to one more track than the \Bu decays,
a single-track reconstruction efficiency uncertainty is assigned to both efficiency ratios.
This 2.3\% uncertainty is not found to depend on \pt or $\abs{y}$~\cite{tracking}.
Several other potential sources of uncertainty were considered and found
to have negligible effects on the efficiency \emph{ratios}.
The muon identification and reconstruction efficiencies, in particular, cancel out.
The efficiencies were also recomputed with varied \Bs \pt distributions
and with the decay angular distributions reweighed
to match the data; both variations have negligible effects.
Finally, the simulated events were reweighed
(with weights dependent on the $y$ and \pt of the \PB meson, as well as on the \pt of the kaons)
so that the \Bs, \Bu, and \Bd MC distributions match the measured ones.
This procedure leads to systematic uncertainties in the 1--2\% and 2--5\% ranges
for the \Rs and \Rd measurements, respectively.

Apart from the uncertainty in the track reconstruction efficiency,
assumed to be independent of \pt and $\abs{y}$,
the bin-to-bin systematic uncertainties are added in quadrature.
For the \Rs measurement, they are in the 2.3--3.2\% and 1.8--4.4\% ranges for
the \pt and $\abs{y}$ results, respectively, while for \Rd the
corresponding ranges are 2.4--7.8\% and 2.3--4.9\%.
The larger \Rd uncertainties arise from the more complex background composition of the \Bd decay.
Both measurements have $\approx$1\% statistical uncertainties in each bin.

\begin{figure}[ht]
\centering
\includegraphics[width=\cmsFigWidth]{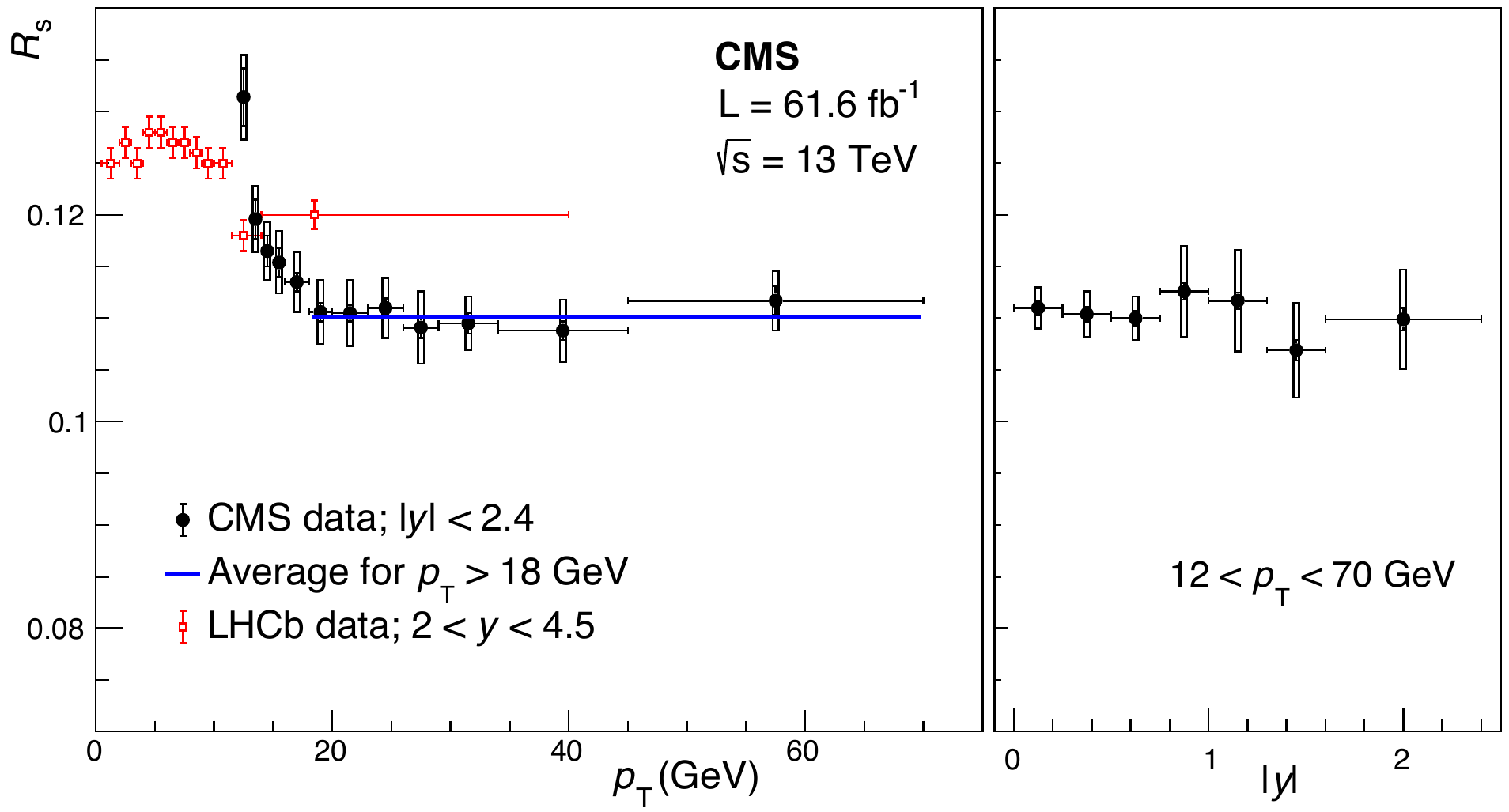}
\caption{Efficiency-corrected yield ratio \Rs,
as a function of \pt (left) and $\abs{y}$ (right).
The vertical bars (boxes) represent the statistical (bin-to-bin systematic) uncertainties,
while the horizontal bars give the bin widths.
The global uncertainty (of 2.3\%) is not graphically represented.
The blue line represents the average for $\pt > 18$\GeV.
For comparison, the LHCb measurement~\cite{1910.09934} is also shown.}
\protect\label{fig:Rs}
\end{figure}

The measured \Rs values, presented in Fig.~\ref{fig:Rs},
do not show any signs of a rapidity dependence.
In contrast, they show a clear \pt-dependence at low \pt,
followed by a flat high-\pt trend.
Averaging the $\pt > 18$\GeV measurements gives $\Rs = 0.1102 \pm 0.0027$,
where the uncertainty includes all contributions, added in quadrature.
The low-\pt dependence is compatible with
the LHCb measurements (for $2<y<4.5$)~\cite{1910.09934}, 
also shown in Fig.~\ref{fig:Rs}.

\begin{figure}[ht]
\centering
\includegraphics[width=\cmsFigWidth]{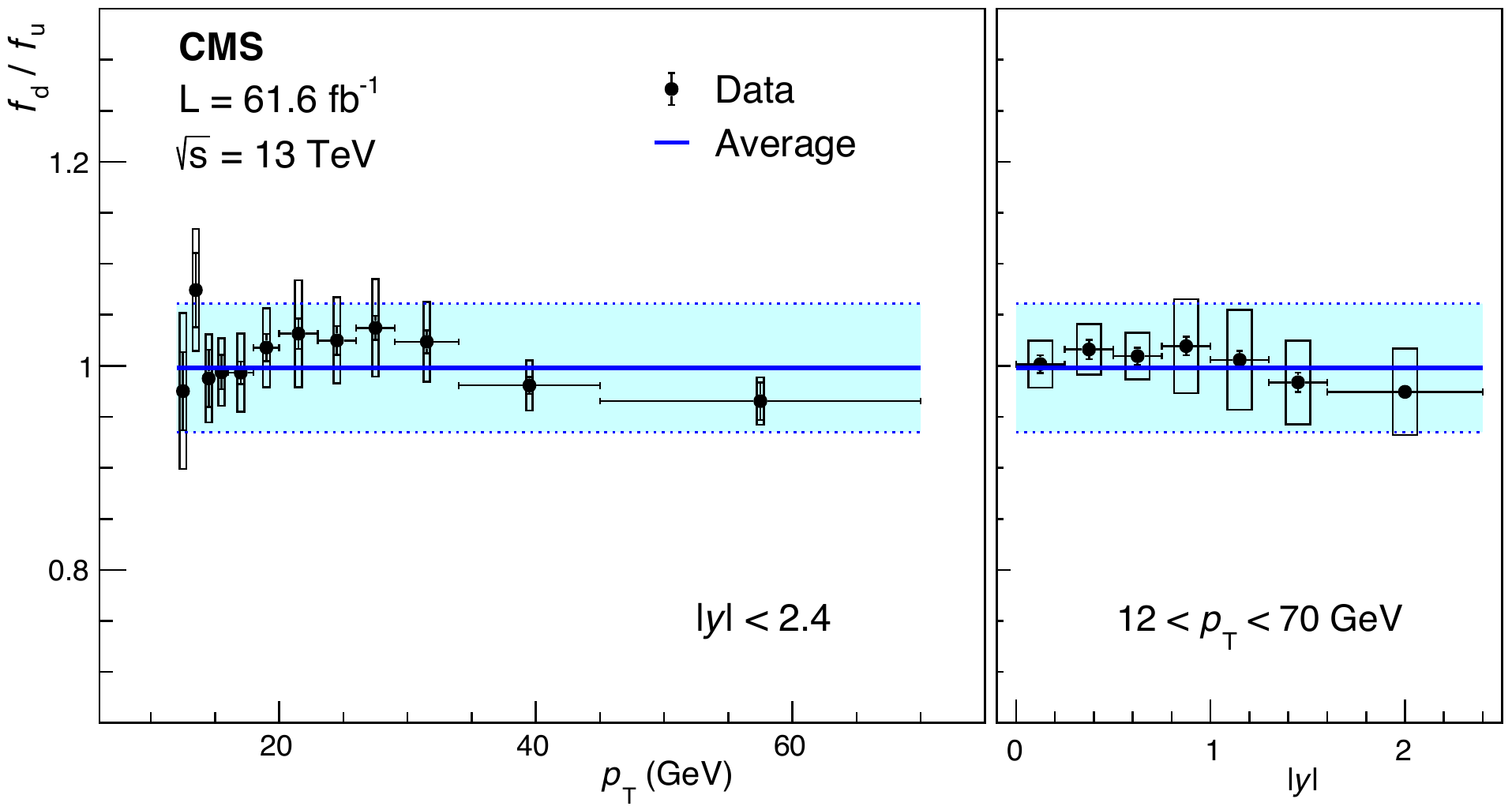}
\caption{The ratio of the \Bd to \Bu hadron production fractions \fdfu,
as a function of \pt (left) and $\abs{y}$ (right).
The vertical bars (boxes) represent the statistical (bin-to-bin systematic) uncertainties,
while the horizontal bars give the bin widths.
The horizontal blue line and band represent the average value and uncertainty. 
The global uncertainty (of 5.7\%) is included in the blue bands but not in the individual data points.}
\protect\label{fig:fdfu}
\end{figure}

The measured \Rd ratio can be used to probe isospin invariance in \PB meson production,
converting it into the \fdfu observable with Eq.~(\ref{eq:fdoverfuDefinition}),
as long as the needed ratio of \PB meson branching fractions,
$\mathcal{B}(\BuJpsiK) / \mathcal{B}(\BdJpsiKstar)$,
is evaluated without the isospin invariance assumption.
For $\mathcal{B}(\BdJpsiKstar)$, we take the world-average value~\cite{PDG2022},
which is dominated by measurements at the $\PGU(4S)$ that assume isospin invariance:
$\Rpmz = \mathcal{B}(\PGU(4S) \to \PBp \PBm) / \mathcal{B}(\PGU(4S) \to \PBz \PABz) = 1$.
For $\mathcal{B}(\BuJpsiK)$, we use its most precise measurement~\cite{BELLE:2019xld},
after correcting for their assumption, $\Rpmz = 1.058 \pm 0.024$~\cite{HFAG2018},
to make it compatible with the branching fractions that use $\Rpmz = 1$.
The ratio of branching fractions
in Eq.~(\ref{eq:fdoverfuDefinition}) is then divided~\cite{Jung_2016}
by the most recent $\Rpmz$ value ($1.059 \pm 0.027$)~\cite{HFAG2022}
to remove the isospin conservation assumption.
The obtained \fdfu ratios are plotted versus \pt and $\abs{y}$ in Fig.~\ref{fig:fdfu}, 
with no dependence on either variable observed.
The average value of $0.998 \pm 0.063$, 
with the uncertainty including all contributions, 
is compatible with unity within the 6\% precision of the measurement, 
consistent with isospin invariance in \PB meson production at hadron colliders.

The numerical results corresponding to Figs.~\ref{fig:Rs} and~\ref{fig:fdfu}
are tabulated in \suppMaterial\ and in the HEPData record for this analysis~\cite{hepdata}.

In summary,
the ratio of the \Bs and \Bu hadron production fractions, \fsfu,
directly proportional to the ratio of the efficiency-corrected meson yields, \Rs,
is studied as a function of the \PB meson transverse momentum \pt and rapidity,
using the \BsJpsiPhi and \BuJpsiK decay channels.
The analysis uses an event sample of $\Pp\Pp$ collisions
at a center-of-mass energy of 13\TeV, collected by CMS in 2018 and
corresponding to an integrated luminosity of 61.6\fbinv.
While no \Rs dependence on the \PB meson rapidity is seen,
a strong variation is observed in the $12 < \pt < 18$\GeV
range, followed by a flat trend for higher \pt values.
The \fdfu ratio, measured 
for the first time in proton-proton collisions 
using the \BdJpsiKstar decay channel,
is found to be compatible with unity
and independent of rapidity and \pt.
This is the first direct measurement of isospin invariance 
in \PB meson production at hadron colliders.
The \mbox{\PQb hadron} production fractions presented in this Letter also
provide a crucial input to measurements by ATLAS and CMS of the $\Bs \to \MM$ branching fraction.

\begin{acknowledgments}
      We congratulate our colleagues in the CERN accelerator departments for the excellent performance of the LHC and thank the technical and administrative staffs at CERN and at other CMS institutes for their contributions to the success of the CMS effort. In addition, we gratefully acknowledge the computing centers and personnel of the Worldwide LHC Computing Grid and other centers for delivering so effectively the computing infrastructure essential to our analyses. Finally, we acknowledge the enduring support for the construction and operation of the LHC, the CMS detector, and the supporting computing infrastructure provided by the following funding agencies: BMBWF and FWF (Austria); FNRS and FWO (Belgium); CNPq, CAPES, FAPERJ, FAPERGS, and FAPESP (Brazil); MES and BNSF (Bulgaria); CERN; CAS, MoST, and NSFC (China); MINCIENCIAS (Colombia); MSES and CSF (Croatia); RIF (Cyprus); SENESCYT (Ecuador); MoER, ERC PUT and ERDF (Estonia); Academy of Finland, MEC, and HIP (Finland); CEA and CNRS/IN2P3 (France); BMBF, DFG, and HGF (Germany); GSRI (Greece); NKFIH (Hungary); DAE and DST (India); IPM (Iran); SFI (Ireland); INFN (Italy); MSIP and NRF (Republic of Korea); MES (Latvia); LAS (Lithuania); MOE and UM (Malaysia); BUAP, CINVESTAV, CONACYT, LNS, SEP, and UASLP-FAI (Mexico); MOS (Montenegro); MBIE (New Zealand); PAEC (Pakistan); MES and NSC (Poland); FCT (Portugal); MESTD (Serbia); MCIN/AEI and PCTI (Spain); MOSTR (Sri Lanka); Swiss Funding Agencies (Switzerland); MST (Taipei); MHESI and NSTDA (Thailand); TUBITAK and TENMAK (Turkey); NASU (Ukraine); STFC (United Kingdom); DOE and NSF (USA).

\end{acknowledgments}
\bibliography{auto_generated}
\ifthenelse{\boolean{cms@external}}{}{
\clearpage
\appendix
\numberwithin{table}{section}
\numberwithin{figure}{section}
\section{Numerical results in bins of \texorpdfstring{\pt and $\abs{y}$}{pt and abs(y)}\label{app:suppMat}}

Tables~\ref{table:Systematics_BspT_hepdata} and~\ref{table:Systematics_Bseta_hepdata} 
provide the numerical \Rs results in bins of \pt and $\abs{y}$, respectively, as shown in\cmsfigone{}. 
Tables~\ref{table:Systematics_BdpT_hepdata} and~\ref{table:Systematics_Bdeta_hepdata} 
provide the numerical \fdfu results in bins of \pt and $\abs{y}$, respectively, as shown in\cmsfigtwo{}. 
Besides the central values, the tables include the statistical and systematic uncertainties for each bin.
Not included in the tables is an additional systematic uncertainty of 2.3\%
associated with the track reconstruction efficiency that applies to all results. 
In addition, the \fdfu results have systematic uncertainties of 4.6\% and 2.5\% 
reflecting the uncertainties in the branching fractions from\cmseqdos{} 
and the \Rpmz correction factor, respectively, that are not included in the tables.

\begin{table}[hbtp!]
\begin{center}
\topcaption{The measured \Rs values as a function of \pt, 
with the statistical ($\sigma_{\text{stat}}$) and bin-to-bin systematic ($\sigma_{\text{sys}}$) uncertainties, in percent.
Not included in the table is an additional systematic uncertainty of 2.3\% 
that is common to all bins and is associated with the track reconstruction efficiency.}
\label{table:Systematics_BspT_hepdata}
\begin{scotch}{cccc}  
\pt  &  \Rs  & $\sigma_{ \text{stat} }$  & $\sigma_{\text{sys}}$ \\
(\GeVns{})   &     & (\%)   & (\%) \\
\hline
12--13 &  0.1314  &  2.1 &  3.1   \\
13--14 &  0.1196  &  1.6 &  2.7   \\
14--15 &  0.1165  &  1.3 &  2.4   \\
15--16 &  0.1154  &  1.2 &  2.6   \\
16--18 &  0.1135  &  0.8 &  2.6   \\
18--20 &  0.1106  &  0.8 &  2.8   \\
20--23 &  0.1105  &  0.7 &  2.9   \\
23--26 &  0.1110  &  0.8 &  2.6   \\
26--29 &  0.1091  &  0.9 &  3.2   \\
29--34 &  0.1095  &  0.9 &  2.3   \\
34--45 &  0.1088  &  0.9 &  2.8   \\
45--70 &  0.1117  &  1.3 &  2.6   \\
\end{scotch}
\end{center}
\end{table}

\begin{table}[hbtp!]
\begin{center}
\topcaption{The measured \Rs values as a function of $\abs{y}$, 
with the statistical ($\sigma_{\text{stat}}$) and bin-to-bin systematic ($\sigma_{\text{sys}}$) uncertainties, in percent.
Not included in the table is an additional systematic uncertainty of 2.3\% 
that is common to all bins and is associated with the track reconstruction efficiency.}
\label{table:Systematics_Bseta_hepdata}
\begin{scotch}{cccc}  
$\abs{y}$  &  \Rs   & $\sigma_{ \text{stat} }$  & $\sigma_{\text{sys}}$ \\
   &     & (\%)   & (\%) \\
\hline
0.00--0.25 &  0.1110  &  0.6 &  1.8   \\
0.25--0.50 &  0.1104  &  0.6 &  2.0   \\
0.50--0.75 &  0.1100  &  0.6 &  1.9   \\
0.75--1.00 &  0.1126  &  0.7 &  3.9   \\
1.00--1.30 &  0.1117  &  0.8 &  4.3   \\
1.30--1.60 &  0.1069  &  1.0 &  4.3   \\
1.60--2.40 &  0.1099  &  1.0 &  4.4   \\
\end{scotch}
\end{center}
\end{table}

\begin{table}[hbtp!]
\begin{center}
\topcaption{The measured \fdfu values as a function of \pt, 
with the statistical ($\sigma_{\text{stat}}$) and bin-to-bin systematic ($\sigma_{\text{sys}}$) uncertainties, in percent.
Not included in the table is an additional systematic uncertainty of 5.7\% 
that is common to all bins and is the sum in quadrature of uncertainties associated 
with the track reconstruction efficiency, 
branching fractions from \cmseqdos, and the \Rpmz correction factor.}
\label{table:Systematics_BdpT_hepdata}
\begin{scotch}{cccc}  
\pt   &  \fdfu    & $\sigma_{ \text{stat} }$  & $\sigma_{\text{sys}}$ \\
(\GeVns{})   &     & (\%)   & (\%) \\
\hline
12--13 &  0.975  &  3.9 &  7.8  \\
13--14 &  1.074  &  3.4 &  5.6  \\
14--15 &  0.988  &  2.8 &  4.4  \\
15--16 &  0.994  &  1.7 &  3.3  \\
16--18 &  0.993  &  1.1 &  3.9  \\
18--20 &  1.018  &  1.3 &  3.8  \\
20--23 &  1.032  &  1.4 &  5.1  \\
23--26 &  1.025  &  1.4 &  4.1  \\
26--29 &  1.037  &  1.1 &  4.6  \\
29--34 &  1.024  &  1.1 &  3.8  \\
34--45 &  0.981  &  0.8 &  2.5  \\
45--70 &  0.965  &  1.9 &  2.4  \\
\end{scotch}
\end{center}
\end{table}

\begin{table}[hbtp!]
\begin{center}
\topcaption{The measured \fdfu values as a function of $\abs{y}$, 
with the statistical ($\sigma_{\text{stat}}$) and bin-to-bin systematic ($\sigma_{\text{sys}}$) uncertainties, in percent.
Not included in the table is an additional systematic uncertainty of 5.7\% 
that is common to all bins and is the sum in quadrature of uncertainties associated 
with the track reconstruction efficiency, 
branching fractions from \cmseqdos, and the \Rpmz correction factor.
}
\label{table:Systematics_Bdeta_hepdata}
\begin{scotch}{cccc}
$\abs{y}$  &  \fdfu  & $\sigma_{ \text{stat} }$  & $\sigma_{\text{sys}}$  \\
   &     & (\%)   & (\%) \\
\hline
0.00--0.25 &  1.002  &  0.9 &  2.3  \\
0.25--0.50 &  1.016  &  0.9 &  2.4  \\
0.50--0.75 &  1.010  &  0.8 &  2.3  \\
0.75--1.00 &  1.019  &  0.9 &  4.5  \\
1.00--1.30 &  1.006  &  0.9 &  4.9  \\
1.30--1.60 &  0.984  &  1.0 &  4.2  \\
1.60--2.40 &  0.975  &  0.6 &  4.3  \\
\end{scotch}
\end{center}
\end{table}

}

\cleardoublepage \section{The CMS Collaboration \label{app:collab}}\begin{sloppypar}\hyphenpenalty=5000\widowpenalty=500\clubpenalty=5000
\cmsinstitute{Yerevan Physics Institute, Yerevan, Armenia}
{\tolerance=6000
A.~Tumasyan\cmsAuthorMark{1}\cmsorcid{0009-0000-0684-6742}
\par}
\cmsinstitute{Institut f\"{u}r Hochenergiephysik, Vienna, Austria}
{\tolerance=6000
W.~Adam\cmsorcid{0000-0001-9099-4341}, J.W.~Andrejkovic, T.~Bergauer\cmsorcid{0000-0002-5786-0293}, S.~Chatterjee\cmsorcid{0000-0003-2660-0349}, K.~Damanakis\cmsorcid{0000-0001-5389-2872}, M.~Dragicevic\cmsorcid{0000-0003-1967-6783}, A.~Escalante~Del~Valle\cmsorcid{0000-0002-9702-6359}, P.S.~Hussain\cmsorcid{0000-0002-4825-5278}, M.~Jeitler\cmsAuthorMark{2}\cmsorcid{0000-0002-5141-9560}, N.~Krammer\cmsorcid{0000-0002-0548-0985}, L.~Lechner\cmsorcid{0000-0002-3065-1141}, D.~Liko\cmsorcid{0000-0002-3380-473X}, I.~Mikulec\cmsorcid{0000-0003-0385-2746}, P.~Paulitsch, J.~Schieck\cmsAuthorMark{2}\cmsorcid{0000-0002-1058-8093}, R.~Sch\"{o}fbeck\cmsorcid{0000-0002-2332-8784}, D.~Schwarz\cmsorcid{0000-0002-3821-7331}, M.~Sonawane\cmsorcid{0000-0003-0510-7010}, S.~Templ\cmsorcid{0000-0003-3137-5692}, W.~Waltenberger\cmsorcid{0000-0002-6215-7228}, C.-E.~Wulz\cmsAuthorMark{2}\cmsorcid{0000-0001-9226-5812}
% ...
\par}
\vskip\cmsinstskip
\dag:~Deceased\\
$^{1}$Also at Yerevan State University, Yerevan, Armenia\\
$^{2}$Also at TU Wien, Vienna, Austria\\
% ...
\end{sloppypar}
\end{document}